\documentclass[review,3p,11.5pt]{elsarticle}  
\usepackage{microtype}
\usepackage [english]{babel}
\usepackage [autostyle, english = american]{csquotes}
\MakeOuterQuote{"}
\usepackage{xcolor}

\usepackage{graphicx}
\usepackage{microtype}  
\usepackage{amssymb}
\usepackage{multirow}
\usepackage{amsmath}
\usepackage{amsfonts}
\usepackage{enumitem}
\usepackage{lscape}
\usepackage{adjustbox}
\usepackage{bm}
\usepackage{subcaption}
\usepackage{stackengine}
\usepackage[mathscr]{euscript}
 \let\mathscr\relax
\usepackage[scr]{rsfso}
\usepackage[ruled,vlined]{algorithm2e}

\journal{NJBDA - JBDTP} 

\begin{document}
\begin{frontmatter}


\title{ 
Strategies for Democratization of Supercomputing: \\
Availability, Accessibility and Usability of High Performance Computing \\
for Education and Practice of Big Data Analytics }


\author[label1]{Jim Samuel}
\author[label2]{Margaret Brennan-Tonetta}
\author[label3]{Yana Samuel}
\author[label4]{Pradeep Subedi}
\author[label5]{Jack Smith.}
\address[label1]{jim@aiknowledgecenter.com, University of Charleston}
\address[label2]{mbrennan@njaes.rutgers.edu, Rutgers University}
\address[label3]{samuel.y@northeastern.edu, Northeastern University}
\address[label4]{pradeep.subedi@rutgers.edu, Rutgers University}
\address[label5]{jack.smith@wvresearch.org, Marshall University}


\begin{abstract}
There has been an increasing interest in and growing need for high performance computing (HPC), popularly known as supercomputing, in domains such as textual analytics, business domains analytics, forecasting and natural language processing (NLP), in addition to the relatively mature supercomputing domains of quantum physics and biology.  
HPC has been widely used in computer science (CS) and other traditionally computation intensive disciplines, but has remained largely siloed away from the vast array of social, behavioral, business and economics disciplines. 
However, with ubiquitous big data, there is a compelling need to make HPC technologically and economically accessible, easy to use, and operationally democratized.  
Therefore, this research focuses on making two key contributions, the first is the articulation of strategies based on availability, accessibility and usability for the demystification and democratization of HPC, based on an analytical review of Caliburn, a notable supercomputer at its inception. The second contribution is a set of principles for HPC adoption based on an experiential narrative of HPC usage for textual analytics and NLP of social media data from a first time user perspective. Both, the HPC usage process and the output of the early stage analytics are summarized.    

This research study synthesizes expert input on HPC democratization strategies, and chronicles the challenges and opportunities from a multidisciplinary perspective, of a case of rapid adoption of supercomputing for textual analytics and NLP. Deductive logic is used to identify strategies which can lead to efficacious engagement, adoption, production and sustained usage for research, teaching, application and innovation by researchers, faculty, professionals and students across a broad range of disciplines. 

\end{abstract}


\end{frontmatter}


\begin{flushleft}
\section{Introduction} \label{S:1}
\begin{center}
\textit{  "Because HPC stands at the forefront of scientific discovery and commercial innovation, it is positioned at the frontier of competition—for nations and their enterprises alike..."     
Ezell \& Atkinson, 2016 } \end{center}
Big data and artificial intelligences (AIs) are having a significant impact on business, work, governance, social interaction and education. While big data and AIs hold tremendous potential for value creation and transformation of human life, its potential can only be realized through appropriate technological implementations. Specifically, significantly more powerful and scaled up data processing, networking and data storage capabilities are required to harness the vast promise of AIs and big data. This necessitates the usage of technologies, which are popularly termed as “Supercomputing”, and also as “High-performance computing” (HPC), referring to the use of supercomputers or high performance computers for computing complex, voluminous or iteration intensive calculations and analytics. The term “supercomputing” (or supercomputers) has been treated as being synonymous with HPC (or high performance computers), and has been parsimoniously described as a computer or a cluster of computers with far greater computing-memory-storage capabilities than a general computer, and as being "\textit{characterized by large amounts of memory and processing power}" \citep{george2020literature}. So also, HPC has been varying defined as being a “\textit{combination of processing capability and storage capacity}” that can efficiently create solutions for “\textit{difficult computational problems across a diverse range of scientific, engineering, and business fields}” \citep{ezell2016vital}, and also as being "\textit{massively parallel processing (MPP) computers}" \citep{bergman2019future}. HPC can be classified as being homogeneous or heterogeneous, based on the use of similar or dissimilar processors (or memory, or similar HPC components) respectively, in its array of processors, such as homogeneous HPC with CPU arrays, and heterogeneous HPC with with CPU and GPU arrays \citep{gao2016survey}. Heterogeneous HPC can be used to improve effectiveness, speed and also to gain additional energy savings. High Performance Computers can therefore be viewed as organized systems of high-powered, parallel structured computational capabilities, including extreme and diverse processing capabilities, general or task varied and scalable memory, scalable storage, grid or network or cloud based, and appropriate capabilities management interfaces with the potential to help solve vast and complex problems. 

\subsection{The Critical Need: Multidisciplinary HPC Applications}
The compelling need for fostering HPC and HPC education has been well recognized by industry, government and academia -however most of these efforts have been siloed, albeit with some progression, to a restricted set of traditionally computational domains. Given the explosive growth in quantity, diversity, complexity, granularity and acceleration of data generation, it has become impossible to meaningfully depend on desktops, servers or standalone computers to create competitive value, and an increasingly large number of disciplines have begun HPC evaluation and adoption processes, to ensure that they remain competitive in progressively data and computation intensive environments \citep{fiore2018road}. The already steep trend towards HPC engagement and adoption can be expected to become stronger with the advent of new technologies and the identification of new opportunities \citep{bergman2019future}. An investigation by the Council on Competitiveness discovered that the vast majority of US corporations with HPC capabilities had significant concerns about being able to hire persons with "sufficient HPC training", and that there are no easy solutions because "\textit{... there aren’t enough faculty, researchers, educators, and professionals with the HPC skills and knowledge to fulfill the demand for talented individuals}" \citep{lathrop2016call}. There has been a sustained call over the past decade for opening up access to HPC / SC resources: "\textit{In the past decade high performance computing has transformed the practice and the productivity of science. Now this analytical power must be opened up to industry, to improve decision making, spur innovation and boost competitiveness}" \citep{Prace_rep}.
HPC has been treated as a critical catalyst for "inter- and trans-discipline breakthroughs" impacting the development of science and innovation globally \citep{mosin2017state}. As illustrations, extant research has called for machine learning applications on varying types of information across a wide range of domains, include behavioral finance, social media analytics, textual data visualization and pandemic sentiment analysis, all of which are best implemented at scale using HPC resources \citep{samuel2017information, rahman2020twitter, Conner2019picture,conner2020picture,samuel2018going}. HPC is now a global phenomena, and competitive advantage in many domains is associated with multidisciplinary HPC capabilities.  

\subsection{The future of HPC impact: Pervaisve \& ubiquitous}
Given the rapid growth of artificial intelligence and the associated need to process big data, HPC has become one of the critical drivers of success for research institutions, corporations and nations. Ubiquitous AI and big data applications imply an equally expansive HPC requirement and influence.  HPC ubiquity is a certainty, though end users will most likely not be required to interface with the technological complexity of HPC systems, just as electricity and internet are ubiquitous, without end users having to interface with systems managing electricity generation equipment or the movement of data packages and transmission protocols for the internet. Institutions, corporations and countries which enable their constituents and stakeholders with easily usable HPC capabilities will possess a significant competitive advantage. HPC driven competitive advantages will impact individuals, businesses, society and nations, bearing the potential for significant socioeconomic impact. It is therefore of paramount importance to look closely at strategies for catalyzing future HPC thought leadership and competitiveness. 


\subsection{The need: Democratized supercomputing!}
Democratized supercomputing, in the context of this study, refers to the opening up of HPC resources, freeing it from restrictive domain boundaries, and making it seamlessly available to all on an as-needed basis. Presently, even where HPC is available, it still remains inaccessible to many. Furthermore, even where access is provided, usability is restricted due to operational and skills barriers.  Our quasi-phenomenological perspective based on the motivations driving the conceptualization, development and deployment of Caliburn supercomputer at Rutgers, the State University of New Jersey, and an analysis of HPC usage for textual analytics and NLP indicate that there are three key strategies that are necessary for democratizing HPC across domains: 1) Availability strategy - HPC resources need to be built and distributed to ensure fair availability, 2) Accessibility strategy - just the mere fact that HPC infrastructure exists in an institution or at a location does not ensure its accessibility, and therefore deliberate steps need to be taken to align and distribute available HPC resources in a manner so as to ensure  HPC accessibility, and 3) Usability strategy - availability and accessibility ensure that end users are empowered to access HPC resources, and yet these alone do not democratize or catalyze HPC utilization without the necessary dimension of ease-of-use. An effective HPC democratization initiative must include the three strategies of HPC availability, HPC accessibility, and HPC usability, to ensure that capabilities are developed to achieve a maximized spectrum of benefits from HPC. \\
\

The rest of this paper is organized as follows: 
First, the study clarifies the multidisciplinary context, provides theoretical lenses from information systems and anchors HPC democratization discussion to the theories of technology adoption and usage. This is followed by an analytical and reflective narrative of the motivations and process for the acquisition and deployment of Caliburn, a supercomputer at Rutgers University. Availability, accessibility, and usability strategies are then elaborated upon in subsequent sections. This is followed by a case analysis of HPC usage for NLP and key principles for sustained usage. The paper concludes with notes on implications, limitations and a motivational conclusion.

\section{HPC engagement, adoption \&  sustained usage - theoretical and applied considerations}

\subsection{The future of HPC relevance: Ubiquitous, multidisciplinary, transdisciplinary \& interdisciplinary}
Extant research meaningfully distinguishes between "multidisciplinarity", "interdisciplinarity" and "transdisciplinarity", wherein "multidisciplinarity draws on knowledge from different disciplines but stays within their boundaries", while interdisciplinarity "analyzes, synthesizes and harmonizes links between disciplines into a coordinated and coherent whole" and transdisciplinarity "integrates the natural, social and health sciences in a humanities context, and transcends their traditional boundaries" \citep{choi2006multidisciplinarity, alvargonzalez2011multidisciplinarity}. We believe that such distinction is valuable. However, since this study does not delve into the nature of disciplinary research, but rather emphasizes the need for HPC to be used across disciplines, simultaneously drawing on knowledge and integrating lessons learnt from the past, irrespective of discipline, hence we employ the word "multidisciplinary" in its broadest sense, inclusive of the properties of interdisciplinarity and transdisciplinarity. 

\subsection{HPC engagement: Theoretical basis}
As with all technologies, there are critical drivers for HPC engagement, adoption and sustained usage. Information Systems (IS) research studies have provided extensive insights into user engagement, adoption \&  sustained usage of technologies. The seminal, and in many senses foundational, technology acceptance model (TAM) theory validated and popularized the concepts of perceived usefulness and ease of use of technologies \citep{davis1989perceived}. Subsequent studies updated TAM, including a broader theoretical basis with a unified perspective leading to the “unified theory of acceptance and  use of technology” (UTAUT), which “highlights the importance of contextual analysis in developing strategies for technology implementation” \cite{venkatesh2003user}. Extant research has also demonstrated that the challenges of technology adoption and usage are subject to information facets, information complexity, equivocality of information and information overload \citep{samuel2016analysis, samuel2017informatics}.  Some technologies, such as blockchains, are pertinent to specific user categories where generally dominant drivers of technology acceptance may have less relevance than factors such as “security, privacy, transparency, trust and traceability aspects” \citep{grover2019perceived}. 

\subsection{HPC usage: From theory to practice}
The generic usability and acceptance model (GUAM), in contrast to UTAUT, provides a significantly better explanation of “behavioral intention (72\%) and technology use (63\%)” for learning innovations \citep{obienu2020user}. This demonstrated that domain or discipline sensitive models have the potential to outperform generic adoption models like TAM or UTAT, due to variations on technological features and characteristics of user groups. Factors such as gender and social characteristics have also been shown to influence user engagement with technologies, such as the indication by prior research that women tend to weigh ease of use more strongly than men who tend to focus on perceived usefulness of the technology \citep{venkatesh2000don}. Supportive technologies, such as anthropomorphic chatbots with human-like natural language processing and communication capabilities, can have a meaningfully positive impact on user perception of the usefulness of technology, and such supportive mechanisms can also support perceived ease of use \citep{rietz2019impact}. Additional theories need to be evaluated to maximize the theoretical basis for HPC and supercomputing engagement. For example, flow theory which refers to the "the holistic sensation that people feel when they act with total involvement" and the state of "flow" where people experience becoming "absorbed in their activity", akin to Chess players and gamers whose intelligences are fully engaged and focused \citep{csikszentmihalyi2014play, csikszentmihalyi1992optimal, koufaris2002applying}. On the applied side, numerous artificial intelligence research initiatives highlight the need for HPC in advancing research in multiple areas, such as natural language generation in the context of social media analytics \citep{garvey2021would}. The present study presents an analysis of a first time users' experience with HPC for textual analytics and NLP, to identify some useful principles that will help potential HPC users to move from theory and strategy to applications and practice.

\section{Caliburn – A story of strategic value creation through ACI availability, accessibility and usability}

The story of Caliburn, the first supercomputer at Rutgers University and in the state of New Jersey, begins in 2011 with the creation of the Rutgers Discovery Informatics Institute (RDI2) by Dr. Manish Parashar, Distinguished Professor, Computer Science. His motivation was to establish a comprehensive and internationally competitive multidisciplinary Computational and Data-enabled Science and Engineering (CDS\&E) institute at Rutgers University that could catalyze and nurture the integration of research and education with advanced computing infrastructure (ACI). Parashar structured RDI2 to provide ACI resources that were available, accessible and usable by offering technologies and expertise to academic researchers and companies that want to take advantage of ACI resources,  but do not have the financial resources or expertise necessary to acquire these human and hardware resources.  It was his vision to one day have a national level ACI at Rutgers University available to all.

\subsection{Sensing the need}
The importance and immediacy of having such ACI competency at Rutgers was further accentuated by the growing role of computation and data in all areas of science, engineering and business, as well as current and future trends in ACI. These included disruptive hardware trends, ever-increasing data volumes, complex application structures and behaviors, and new first-order concerns such as fault-tolerance and energy efficiency. These trends are a result of the continued quest towards extreme scales in computing and data that is necessary to drive innovations in science, engineering and other data intensive fields. 

\subsection{Innovation}
The CI developed by RDI2 is innovative and provides researchers with global linkages to the national and international CI (e.g., XSEDE, OSG, OOI, LHC, iPlant, PRACE, EGI, etc.) that connects Rutgers with observational instruments, data streams, experimental tools, simulation systems and individuals distributed across the globe. Overall, the impact of RDI2 is a revolutionary advance in the scale and effectiveness of science and engineering research conducted at Rutgers and by academia and industry throughout the state. 

\subsection{Strategic Vision}
As a next step, it was critical that Rutgers develop and implement a bold strategic vision for an ACI ecosystem that was competitive at the national and international levels \citep{Caliburn1}. This ecosystem had to provide researchers with cutting-edge computing and data handling capabilities, and students with necessary ACI exposure and training. In 2013, RDI2 initiated a university-wide ACI strategic planning process with input from faculty across many disciplines at Rutgers. This resulted in a comprehensive plan, “Accelerating Innovation Through Advanced Cyberinfrastructure: A Strategic Vision for Research Cyberinfrastructure at Rutgers” \citep{Caliburn2}. The plan called for strategic investment in ACI to drive innovation, improve research capabilities and productivity, and enhance faculty competitiveness. Two specific findings of the Rutgers ACI strategic plan were the need to deploy a nationally competitive advanced cyberinfrastructure and to establish a central Office of Advanced Research Computing at the University. 

\subsection{Availability}
Deploying a nationally competitive ACI required infrastructure investments in computing, mass storage, and high speed/bandwidth digital communication that could provide state-of-the-art capacities and capabilities for Rutgers researchers and offer a competitive advantage among Big Ten peer institutions and beyond. However, the level of investment required to achieve this goal was significant. Fortunately, in 2013 the State of New Jersey announced the Higher Education Equipment Leasing Fund to support investments in cutting-edge equipment at the state’s higher education institutions.  RDI2 submitted a proposal entitled “Rutgers University Advanced Compute \& Data Cloud” to establish a statewide ACI resource at Rutgers that could have far-reaching benefits for higher education institutions, industry and state government. The state recognized the tremendous impact that this capability could have for its innovation economy, and the proposal was approved.  RDI2 was awarded \$10 million to purchase cutting edge ACI systems, the largest award given through this program. 

\subsection{Accessibility}
After 2 years of design and installation, Caliburn and companion system ELF were deployed in 2016 at Rutgers. The rationale for the two systems was recognition of the potential limitations for many researchers who are inexperienced in a complex ACI such as Caliburn, or do not need its high-level capabilities. Thus ELF, which had significantly more computing capability than what was currently available but not as complex or powerful as Caliburn, could be used by researchers as a first step in building experience and understanding of more sophisticated ACI. These systems provide a balanced advanced computational and data environment that contains a large-scale high-end compute engine, as well as significant co-located storage with embedded analytics capabilities. The Caliburn and ELF platforms are seamlessly accessible as a cloud service, providing researchers, students, industry and government across the state, with on-demand and pervasive access to these capabilities for research and instruction. Caliburn/ELF are also connected with high-speed networking to key national and international research/educational facilities. The overall platform is unique, and the most powerful academic system in the state. When it was commissioned in Summer 2016, Caliburn was ranked on the Top500 list of computer systems worldwide as \#\\2 system among the US Big 10 institutions and \#\\8 among all US academic institutions, \#\\50 among academic institutions globally, and \#\\166 among all computer systems worldwide \citep{Caliburn3}.

\subsection{Usability}
To further have ACI as available, accessible and usable as possible at the University, RDI2 began planning the creation of the first central Rutgers Office of Advanced Research Computing (OARC) in 2015.  In Spring 2016, the university’s first Associate Vice President for Advanced Research Computing was appointed to continue the growth and development of university-wide ACI. The functions of this office are to provide strategic leadership, coordinate investments in ACI and related expertise, and catalyze and nurture cyberinfrastructure-enabled multidisciplinary research, all aimed at fostering a community of excellence in computating and data, empowering research, learning, and societal engagement and providing a competitive advantage to the Rutgers community and throughout the state and region. In 2019, management of the Caliburn and ELF systems was moved to OARC. These and other ACI resources are broadly available through this office. 

\subsection{Caliburn: Extending availability, accessibility \& usability with a futuristic vision}
While Caliburn has been used for a wide range of multidisciplinary projects, one of the noteworthy initiatives which embodies the availability - accessibility - usability paradigm is the "Caliburn Supercomputing Awards". This initiative provides a pathway for scholars and academics outside of Rutgers to apply for HPC allocations, and thus expands the reach and impact of Caliburn - this is an example of a useful mechanism for the democratization of HPC: Scholars and academics from institutions without HPC capabilities are now empowered in their research and thought leadership, which would otherwise be lacking. The process which was established by RDI2 and continues with OARC, starts with OARC inviting proposals for "the allocation of computing resources on Caliburn", providing "high-performance computing capabilities to academic researchers across the state to accelerate research programs that use or develop highly scalable computing applications". OARC also invites applicants in a second "startup" category and these startup proposals "are provided as means to have full access with a limited time usage allocation", and are encouraged to be structured such that "they can be converted into awarded allocations during the next call for proposal cycle". Applications are limited to academic institutions in New Jersey under this program - this is something that OARC can go beyond, subject to availability of resources after meeting state-level needs, to provide a measure of access to meet the HPC needs of individual residents of NJ, non-academic institutions and academic outside of the state as well. For example, it would be of great value to NJ residents and students, if public libraries in NJ were empowered to provide interactive HPC demonstrations and interaction opportunities locally.  Another avenue would be to explore relationships with specific institutions outside of the state to foster democratization of HPC resources.

 \section{The Expanding HPC Landscape: Notable initiatives and case analyses}
 There are numerous existing efforts which in some form address the Availability, accessibility and usability paradigm. However, the focus Is mostly on facilitating availability, and accessibility to a lesser extent. Usability tends to be left for the end user to wrestle with, with the help of “user guides”, often leading to a loss of time and effort. The section below provides a brief overview illustrating some of the prominent HPC efforts. 

\subsection{HPC across institutions and disciplines}
Advanced Cyberinfrastructure Research \& Education Facilitators (ACI-REF): The Clemson-led ACI-REF program (NSF \#\\1341935) advanced research computing through a network of Cyberinfrastructure (CI) Facilitators. This team is now co-leading Campus Research Computing Consortium (CaRCC). CaRCC is a NSF Research Coordination Network (NSF \#\\1620695) and a follow-on to the ACI-REF project that addresses the huge growth in demand for local research computing, by sharing, collaborating, and developing best practices for research-facing, system-facing, software-facing, and stakeholder-facing CI professionals. CaRCC doesn’t do large-scale workforce development for CI professionals itself, but most CaRCC institutions have now participated in Neeman’s Virtual Residency Program.  The ACI-REF Virtual Residency Program (VRP) was initiated by Henry Neeman, University of Oklahoma. He was a collaborator on the original ACI-REF proposal, and started the ACI-REF Virtual Residency Program (VRP) with an NSF CC-IEE grant (NSF \#\\1440783) to provide national-scale CI Facilitator training. All the original ACI-REF institutions have participated in the VRP. [Subsection reference: \cite{neeman2016advanced, neeman2018progress}].

\subsection{Facilitation of CI Driven Research}
XSEDE Campus Champions (CCs): There are more than seven hundred Campus Champions, and these numbers are growing, at over three hundred US institutions helping their local researchers use CI, especially large scale/advanced computing. Most CCs perform CI facilitation activities, and CCs usually peer-mentor each other. The Champion community has: (a) a very active mailing list, where CCs exchange ideas and help each other solve problems; (b) regular conference calls for learning what's happening both among CCs and in national CI; and (c) major participation at national conferences like PEARC (e.g., 23\% of PEARC’20 attendees were CCs). Many CCs have also participated in the VRP. The Society of Research Software Engineering (SRSE) has 29 participating universities, supports Research Software Engineers (RSEs), focusing on reproducibility, reusability, and accuracy. The goal is to foster career paths for academic RSEs and ensure that they are recognized and rewarded. The United States Research Software Engineer Association (US-RSE) is the US counterpart for SRSE. It has over 700 members. The US Research Software Sustainability Institute (URSSI) has been funded by NSF from 2017-21. Its goal is to design an institute on research software and to build the RSE community, in order to (i) improve how individuals and teams function, and (ii) advance research software and the STEM research it supports. Other efforts include initiatives such as the Supercomputing in Plain English (SIPE) workshop, which is an annual workshop on supercomputing (HPC) at Oklahoma University, run by Henry Neeman, using plain English to introduce fundamental issues of supercomputing as they relate to Computational and Data-enabled Science \& Engineering. Internet2 and EDUCAUSE also have programs to help enable CI Facilitators [Subsection reference: \cite{neeman2016advanced, neeman2018progress}].

 \subsection{NLP Case Analysis: Multidisciplinary HPC education \& productivity principles}
 In addition to the analysis and study of institutional level initiatives, it is important to factor in individual perspectives of HPC usage from a multidisciplinary perspective. This subsection summarizes the key conceptual factors and implications for HPC usage from the lens of a first time HPC user for textual analytics (TAn) and NLP on social media data. TAn and NLP have been widely used for social media data analytics and a broad range of natural language sense-making efforts, including research on COVID-19, stock market and public perception \citep{samuel2020covid,kretinin2018going, samuel2020feeling}. The narrative is based on a Caliburn allocation award for a TAn \& NLP project - this narrative does not focus on the findings of the core research, but rather on the process employed by a new HPC user, and the associated learning curve. The analysis highlights how the Caliburn award served as an excellent example of CI availability and facilitation of capability, but was lacking in sufficient usability support for a non-CS user. 
 
 \subsubsection{Caliburn usage case: HPC for making sense of TAn and NLP}
The goals \& motivation for HPC engagement were to make sense of a large data file of social media data, consisting of over seven million records, which needed to cleaned, explored, summarized, and analyzed for general and public sentiment insights. The analysis was initiated using R and Python, and associated software packages and libraries. This task, which was initiated in 2019 was beyond the capabilities of a high powered desktop with 64 GB of RAM, and mandated the use of HPC resources. A Caliburn HPC allocation award made this analysis possible from 2020, and the second phase of the project continues into 2021. The sentiment analysis and custom advanced data visualization methods used for analysis necessitated the installation of new packages and libraries on the HPC system. 

\subsubsection{Navigating Caliburn: Initiation}
The HPC engagement process involved remote access of Caliburn with a two factor authentication process. Rutgers OARC had a very clear step-wise process for this, and the initial access process was smooth and efficient, thus indicating that the availability of CI HPC resources were well supported by an efficient accessibility strategy. The challenges occurred on the usability level, and though issues such as data transfer were self-resolved by the researcher's own efforts, issues with running necessary software remained. The major challenges faced were on key dimensions of usability: a) interface - the command line drive interface led to a long learning curve, and this could be mitigated by the use of open source solution such as OpenOndemand, b) software installation - while Caliburn had existing tools, it was a laborious and iterative process to figure out optimal ways to install all required R packages and Python libraries, and c) exporting and saving the analysis, especially the data visualizations in the required format. A qualitative estimate indicates that about 75 \% of man-hours invested into the analysis were spent of resolving usability issues - this indicates a significant challenge for new users and non-CS users of HPC. Two kinds of HPC engagement were utilized: running live jobs (smaller subsets of data, relatively low computational requirement) and running batch jobs (the intended purpose of HPC -using larger datasets, with higher computational requirements). In running batch jobs, an additional issue arose: development of standalone script for running complete analytical processes - this requires a new mindset as compared to live-interactive analytics processes, and is described in further detail in the Caliburn usage process sub-section below. 

\subsubsection{Caliburn usage process: Tactical summary}
In our case, we utilized two kinds of HPC engagement: running live jobs and running batch jobs. Running live jobs involved “asking” Caliburn for access to a HPC node, where once access was provided, it was like using a Linux machine via terminal. This use case is particularly suitable for analyses that require frequent user-intervention and employs smaller subsets of data with relatively low computational requirements. In contrast to live jobs, running batch jobs can be further divided into small scale jobs and large-scale jobs. Both batch cases require the development of a standalone script for running complete end-to-end analytical processes - this requires a new mindset as compared to live-interactive analytics processes. While the small-scale batch job is similar to running a live job from the perspective of storage and computational requirement, the benefit is that once the script is written, it does not require constant monitoring and interaction. Large-scale batch jobs are mainly designed for larger datasets with higher computational requirements, and it usually engages multiple nodes of the HPC machine. However, these large-scale batch jobs come with a steep learning curve on the efficient use of multiple nodes and cores of a HPC machine, and on end-to-end output-inclusive scripts that are required to run parallel analytics processes. 

\subsubsection{Caliburn usage case: Deductive implications}
The above described HPC usage analysis for TAn and NLP provided interesting principles and insights for HPC adoption: 1) Effective HPC availability and accessibility strategies led to the use of Caliburn for business analytics purposes, employing TAn and NLP methods. This research would not have been possible with the Rutgers-OARC's Caliburn award; 2) Usability strategies were limited to benefit expert HPC users and were not ready to support non-CS and new HPC users; 3) OARC support staff were well trained and provided critical assistance in addressing the barriers and issues with data transfers, software installation and script required to enable TAn and NLP tools - this mitigated usability issues but still led to a significant increase in man-hours used for the analysis; 4) The absence of a GUI (graphic user interface) and user friendly software installation process caused delays and discouraged additional creative research efforts, thus limiting the value created through for core research objectives; and 5) a successful HPC usability strategy must implement appropriate user education and critical changes to the HPC process and system, to enable smooth and high impact multidisciplinary research.     

\subsection{HPC education \& productivity principles}
User education is critical for the success of any HPC usability strategy. Education in High Performance Computing is an actively developing field with specialists from various disciplines participating in diverse initiatives. The development of specialized and purpose specific educational programs, is therefore critical for meeting the ever-increasing demand for skilled HPC users \citep{connor2016next}. It has been shown that addressing the skill set gap is critical for meeting the needs in CI research workforce as well and interdisciplinary teaming helps foster the learning process in technological domains \citep{choi2017vertically}. Extant research has shown that interdisciplinary faculty are essential for successful implementation of HPC instruction \citep{neumann2017interdisciplinary}. Globally, emphasis on hands-on experiences and communications with international faculty is gaining significant prominence in HPC education \citep{sancho2016bsc}. Several researchers call for a “holistic approach” to HPC training and education rather than focusing on a particular HPC ecosystem \citep{chaudhury2018let}. CI productivity can be maximized with an effective implementation of HPC usability strategy. 











\section{HPC democratization strategies: Availability, accessibility and usability}
This leads us to summarizing a key contribution of this study:  the articulation of  availability, accessibility and usability strategies for the demystification and democratization of HPC. 
\subsection{Availability}
This has been the first step in expanding HPC usage, and institutions across the nation and globally have been at the forefront of acquiring and developing supercomputing capabilities. An effective availability strategy consists of acquiring and developing CI, such that it caters to stakeholder needs for the current phase, while being scalable to accommodate larger workloads, and flexible to be developed for diverse workloads. In the age of cloud computing, needless to say, availability is not restricted by geography but bounded by network and access protocols. Once capacities were developed across many institutions, it was observed that under-utilization was a problem due to accessibility issues. Multiples measures exist for the scale and scope of CI, which are essentially a summary of HPC technological components such as processors, memory, storage and structure.  

\subsection{Accessibility}
While some CI is built to cater to a very limited and specialized group of stakeholders, the accessibility strategy is defined in reference to HPC capabilities at public organizations and academic institutions, where there is a need to cater to a broader need for HPC resources, as well as a responsibility to maximize the investment dollars.  An effective accessibility strategy consists of frameworks and processes that maximize the engagement of multidisciplinary stakeholders, plan for expanded user categories with prioritization of core stakeholders, such that under-utilization of allocation is minimized. Accessibility strategies should be augmented with appropriate tracking and reporting of CI productivity and reach to evaluate the success of availability strategy. The minimization of under-utilization of allocation of resources would serve as an indicator of success of an accessibility strategy. 

\subsection{Usability}
There is a difference between optimal allocation of CI resources and optimal usage of CI resources. Maximizing the allocation of resources would be a measure for effectiveness of an accessibility strategy, but that does not ensure optimal utilization and productivity at the end user level. Productivity maximization at the end user level not only requires good availability and accessibility strategies, but also a robust usability strategy. 
Based on lessons learnt from Caliburn, usage experiences, technology engagement theories and a broader HPC landscape review, we describe a usability strategy to maximize end-user level productivity. An effective HPC usability strategy consists of a well designed HPC system with multidisciplinary orientation, easy to use human interfaces, expert usage support, and general and discipline specific applied HPC education. Productivity maximization at the end user level would serve as an indicator of success for usability strategies. Basic measures, for example, could use averages of ratios of actual consumption to total allocation per user, across users in a category. Such ratios and measures would also serve as a check on over-allocation ( excess availability at an individual end-user level) as well. An effective usability strategy has been missing across many CI initiatives and addressing this strategy can lead to a remarkable increase in HPC productivity without additional expensive investments into increasing HPC availability. 

\section{Implications}
HPC as an evolving technological, economical and social multidisciplinary paradigm will have significant implication for human society, and this topic by itself will require a fair amount of research. This section does not attempt to discuss all potential implications, but is restricted to select issues most relevant to the current narrative - namely, the use of HPC productivity optimization strategies, HPC education as a key to HPC democratization and special issues and equal opportunity concerns with multidisciplinary HPC.   

\subsection{The future of HPC expansion: Decreasing depth \& increasing breadth}
Moore’s law predicted the doubling of transistors every two years, and the processor industry has experienced this curve till it reached the limits of physical properties – therefore, Moore’s law will no longer be relevant to the future of new technologies, to the same extent that it has been in the past \citep{moore1965cramming,  moore1995lithography, schaller1997moore, theis2017end}. HPC has entered into the early stages of a post-Moore era, and we have also seen significant advances in forms of massively parallel and hybrid forms of scalable computing. The underlying technologies have proven their worth and are well understood, leading to the greatest need: satisfy a broad range of domain specific big data and voluminous algorithmic processing. We posit that its future is going to be relatively more strongly driven by an expansion the breadth of HPC applications across domains, rather than intensifying the vertical implementation of hardware improvements for marginal benefits in size and speed. This is supported by a growing demand for newer HPC applications, and relatively dwarfed demand for more sophisticated hardware. The success of cloud based computing services such as Amazon Web Services (AWS) attest this perspective on current trends. Therefore, HPC availability and usability strategies must be revised to cater to a broad range of disciplines, many of which will be traditionally non-computation-intensive disciplines. Development of HPC usability strategies, and HPC education modules in  particular, will have a significant impact on HPC democratization and productivity.   
\subsection{The future of HPC education: Modular, applied and multidisciplinary }
The goal here is to very briefly provide an impetus for HPC education - mostly focused on multidisciplinary curricular "modularization" and democratization. Notable initiatives include the "CyberAmbassadors" program, which is a 2017 CyberTraining project, focused on developing an open source curriculum on interpersonal communication and mentoring skills for CI professionals. Similarly, SIGHPC Education Chapter is a virtual chapter of ACM’s Special Interest Group in HPC and has merged with the IHPCTC (International HPC Training Consortia). They focus on developing best practices for HPC training, but don’t provide such training themselves. The Linux Clusters Institute (LCI) holds workshops on HPC system administration, at introductory, intermediate and advanced levels. These workshops have been extremely successful, typically attracting 20-40 HPC system administrators per workshop. LCI focuses on system-facing CI professionals, not researcher-facing. The Carpentries is an international volunteer organization that has run 2300+ hands-on workshops on research computing skills that so far have served 56,000+ researchers at 250+ institutions worldwide. They have demonstrated the effectiveness of “training the trainers” of researchers in informal education at large scale, with an emphasis on technical skills and pedagogy, and not on training CI Facilitators. The Coalition for Academic Scientific Computation (CASC): Members of this non-profit organization are primarily US institutional CI leaders, plus some national CI leaders. CASC’s focus closely aligns with CI leadership. CASC isn’t budgeted for or positioned to take on a major teaching or training role, but significant peer mentoring emerges from CASC activities. Science Gateways Community Institute (SGCI): The SGCI offers workforce development via internships, mentoring and travel funding to conferences for graduate and undergraduate students, connections to the Young Professional Network, and support for gateway-related career paths [Subsection reference: \cite{neeman2016advanced, neeman2018progress}]. In spite of many such initiatives, there is a critical need to evaluate usability strategy oriented education as a number of these initiatives cater to education pertaining to availability and accessibility, leaving a huge gap in education needs for implementing effective HPC usability strategies.

\subsection{HPC education and democratization: Special issues and equal opportunity}
Some of the challenges in the practice of HPC will be associated with addressing bias, such as gender bias in technology resulting in a low representation of women in HPC practice. Although there is very limited research done on the issue of gender and HPC, it is safe to assume that attracting and retaining women in HPC practice is going to be a challenge. This is already so for the fields of STEM and especially Computer Science, which is affected the most by a significant under-representation of women \citep{ehrlinger2018gender}. Studies conducted on the topic of female under-representation found that less than 20\% of the technological workforce is estimated to be women” \citep{frachtenberg2020representation}. The adverse implications of having so few women in HPC are numerous and significant \citep{frantzana2019women}. Extant research has emphasized the need to develop gender specific learning strategies which accommodate women learners in technology disciplines \citep{samuel2020beyond}. These issues will need to be addressed as aspects of the usability strategy to ensure fair and balanced HPC democratization without bias, facilitating equal opportunities to persons in all categories. 





\section{Conclusion}
This study identifies three key strategies for HPC democratization and important principles for catalyzing multidisciplinary HPC productivity. This research thus provides critical ideas and motivations for promoting HPC based research, applications and innovation in traditionally non-CS and non-computation-intensive disciplines and domains. We believe that this is a vital need for the current decade, and anticipate that this study will contribute to the body of knowledge that will influence HPC education policy in the future. 
We boldly propose and call for an increased emphasis on accessibility and usability strategies: every institution of higher education must ensure some measure of access to HPC resources through partnerships and networks, such as XSEDE, for their faculty and students. The responsibility for such efforts must be shared between those who "own" CI resources and those who need it. \\

\begin{center}
\textit{"HPC is becoming a major driver for innovation offering possibilities that currently we cannot even evaluate or think about" \citep{puertas2020high}} \end{center}

We have emphasized that the dimension which needs the most attention is multidisciplinary HPC education within the HPC usability strategy. We posit that undergraduate and graduate programs across disciplines must contain courses with HPC concepts and application modules, such as HPC lessons in information systems courses for undergraduate business programs and in MIS courses for graduate business and relevant MS programs. 
Furthermore, workshops and interactive virtual education modules can be used for topic and discipline specific training. An appropriate HPC-Usability strategy and forward looking HPC education modules will ensure demystification of HPC and popularize its usage for innovation and value creation, by a broader range of students from multiple disciplines, and thus nurture the future HPC and AI workforce. 
 Furthermore, it is obvious that institutions and corporations with fewer resources, lacking research and technological infrastructure development funding  are at a disadvantage when it comes to HPC usage for research, teaching and practice. An invigorated vision to democratize HPC reach into the smallest of institutions, going beyond boxed-in notions of traditionally bounded HPC domains,  will maximize the return on investment for CI resources, as well as promote the noble and forward looking cause of better educating a robust future technological workforce. 

\end{flushleft}







\bibliography{icimbib}
\bibliographystyle{mod2}\biboptions{authoryear}








\end{document}